\documentclass[11pt]{article}
\usepackage{epsfig}
 
\def\be{\begin{equation}}
\def\ee{\end{equation}}

\def\x{{\bf x}}       
\def\y{{\bf y}}
\begin{document}                             

\title{Quantum Clustering}
\author{David Horn and Assaf Gottlieb\\
School of Physics and Astronomy \\
Raymond and Beverly Sackler Faculty of Exact Sciences \\
Tel Aviv University, Tel Aviv 69978, Israel}
\date{\today}
\maketitle

\begin{abstract}
  We propose a novel clustering method that is based on physical
intuition derived from quantum mechanics. Starting with given data points,
we construct a scale-space probability function. Viewing the latter
as the lowest eigenstate of a Schr\"odinger equation,
we use simple analytic operations
to derive a potential function whose minima determine cluster centers.
The method has one parameter, determining the scale over
which cluster structures are searched.
We demonstrate it on data analyzed in two dimensions (chosen from the
eigenvectors of the correlation matrix).
The method is applicable in higher dimensions by limiting
the evaluation of the Schr\"odinger potential to the locations of
 data points. In this case the method may be formulated in terms
of distances between data points.
\end{abstract}

\section{Introduction}
  Clustering of data is a well-known problem of pattern recognition,
covered in textbooks such as \cite{jain,fukunaga,duda}.
The problem we are looking at is defining clusters
of data solely by the proximity of data points to one another.
This problem is one of unsupervised learning, and is in general
ill defined. Solutions to such problems can be based on intuition
derived from Physics. A good example of the latter is the algorithm
by \cite{bla-96} that is based on associating points with Potts-spins
and formulating an appropriate model of statistical mechanics.
We propose an alternative that is also based on physical intuition,
this one being derived from quantum mechanics.

   As an introduction to our approach we start with the scale-space
algorithm by \cite{roberts} who uses a Parzen-window estimator\cite{duda}
of the probability distribution leading to the data at hand.
The estimator is constructed by associating a Gaussian with each
of the $N$ data points in a Euclidean space of dimension $d$
and summing over all of them. This can be represented, up to
an overall normalization by
\be
\psi(\x) =  \sum_i e^{-{({\bf x -x}_i)^2 \over 2\sigma^2}}
\ee
where $\x_i$ are the data points.
Roberts \cite{roberts} views the maxima  of this
function as determining the locations of cluster centers.

 An alternative, and somewhat related method, is
 Support Vector Clustering
(SVC) \cite{nips00} that is based on a Hilbert-space analysis.
In SVC one defines a transformation from data space
to vectors in an abstract Hilbert space.
SVC proceeds to search for the minimal sphere surrounding these states
in Hilbert space.We will also
associate data-points with states in Hilbert space. Such states
may be represented by Gaussian
wave-functions, in which case
$\psi(\x)$ is the sum of all these states.
This is the starting point of our method.
 We will search for the
Schr\"odinger potential for which $\psi(\x)$ is a ground-state.
The minima of the potential define our cluster centers.

\section{ The Schr\"odinger Potential}

We wish to view $\psi$ as an eigenstate
of the Schr\"odinger equation
\be \label{sch}
H\psi \equiv ( -{\sigma^2 \over 2}\nabla ^2 + V({\bf x}))\psi = E \psi.
\ee
Here we rescaled $H$ and $V$ of the conventional quantum mechanical
equation
to leave only one free parameter, $\sigma$.
For comparison, the case of a single point at $\x_1$ corresponds to
Eq. \ref{sch} with $V={1 \over 2\sigma^2}({\bf x-x}_1)^2$ and $E=d/2$,
thus coinciding with the ground state of the harmonic oscillator
in quantum mechanics.

    Given $\psi$ for any set of data points we can solve Eq. \ref{sch} for $V$:
\be
 V({\bf x})=E +{{\sigma^2 \over 2}\nabla ^2 \psi \over \psi}
 = E - {d \over 2} +{1 \over 2\sigma^2 \psi}\sum_i (\x-\x_i)^2
e^{-{({\bf x -x}_i)^2\over 2 \sigma^2}}
\ee
Let us furthermore require that min$V$=0. This sets the value of
\be
E= - {\rm min} {{\sigma^2 \over 2}\nabla ^2 \psi \over \psi}
\ee
and determines $V(\x)$ uniquely.
$E$ has to be positive since $V$ is a non-negative function. Moreover,
since the last term in Eq. 3 is positive definite, it follows that
\be
0<E\leq {d \over 2} \, .
\ee
We note that $\psi$ is positive-definite. Hence, being an eigenfunction
of the operator $H$ in Eq. (\ref{sch}), its eigenvalue $E$
is the lowest eigenvalue of $H$,
i.e. it describes the ground-state.
 All higher eigenfunctions have nodes whose number increases
as their energy eigenvalues increase.\footnote{In quantum mechanics, where
one interprets $|\psi|^2$ as the probability distribution,
all eigenfunctions of $H$
have physical meaning. Although this approach could be adopted, we have
chosen $\psi$ as the probability distribution
because of simplicity of algebraic manipulations.}

 Given a set of points defined within some region of space, we expect $V(\x)$ to
grow quadratically outside this region,
and to exhibit one or several local minima within the region.
Just as in the harmonic potential of the single point problem,
we expect the ground-state
wave-function to concentrate around the minima of the potential $V(\x)$.
Therefore we will identify these minima with cluster centers.

As an example we display results for the
crab data set taken from Ripley's book \cite{ripley}. These data, given
in a five-dimensional parameter space, show nice separation of the
four classes contained in them when displayed in two dimensions spanned
by the 2nd and 3rd principal components\cite{pca} (eigenvectors) of
the correlation matrix of the data.
The information supplied to the clustering
algorithm contains only the coordinates of the data-points. We display
the correct classification to allow for visual comparison of the clustering
method with the data.
Starting with $\sigma=1/\sqrt 2$ we see in Fig. 1 that the Parzen probability
distribution, or the wave-function $\psi$, has only a single maximum.
Nonetheless, the potential, displayed in Fig. 2, shows already four minima
at the relevant locations.
 The overlap of
the topographic map of the potential with the true classification
is quite amazing.
The minima are the centers of attraction of
the potential, and they are clearly evident although
the wave-function does not display local maxima at these points.
The fact that $V(\x)=E$ lies above the range where all valleys merge
explains why $\psi(\x)$ is smoothly
distributed over the whole domain.

 We ran our method also on the iris data set \cite{iris},
 which is a standard benchmark
obtainable from the UCI repository \cite{UCI}.
The data set contains 150 instances each composed of four
measurements of an iris flower.
There are three types of flowers, represented by 50 instances each.
Clustering of these data in the space of the first two principal components
is depicted in Figure 3. The results shown here are for $\sigma=1/4$
which allowed for clustering that amounts to misclassification of three
instances only, quite a remarkable achievement (see, e.g. \cite{nips00}).

\section{ Application of Quantum Clustering}

The examples displayed in the previous section show that, if the
spatial representation of the data  allows for meaningful clustering
using geometric information, quantum clustering (QC) will do the job.
There remain, however, several technical questions to be answered:
 What is the preferred choice of
$\sigma$? How can
QC be applied in high dimensions?
 How does one choose the appropriate space, or metric, in
which to perform the analysis?
We will confront these issues in the following.

  In the crabs-data of we find
that as
$\sigma$ is decreased to ${1 \over 2}$, the previous minima of $V(\x)$
get deeper
and two new minima are formed. However the latter are insignificant, in
the sense that they lie at high values (of order $E$), as shown in Fig. 4.
Thus, if we classify data-points to clusters according to their topographic
location on the surface of $V(\x)$, roughly the same clustering assignment
is expected for  $\sigma = {1 \over 2}$ as for ${1 \over \sqrt 2}$.
One important advantage of quantum clustering is that $E$ sets the
scale on which minima
are observed. Thus, we learn from Fig. 2 that the cores of all four clusters
can be found at $V$ values below $.4E$. The same holds for the
significant minima of Fig. 4.

 By the way,
the wave function acquires only one additional
maximum at $\sigma={1 \over 2}$.
As $\sigma$ is being further decreased,
more and more maxima are expected in $\psi$
and an ever increasing number of minima (limited by $N$) in $V$.

The one parameter of our problem,
$\sigma$, signifies the distance that we probe.
Accordingly we expect to find clusters relevant to proximity information
of the same order of magnitude.
One may therefore vary  $\sigma$ continuously and look for stability
of cluster solutions, or limit oneself to relatively low values of
 $\sigma$ and decide to stop the search once a few clusters
 are being uncovered.

\section { Principal Component Metrics}

In all our examples data were given in some high-dimensional space and
  we have analyzed them after defining a projection and a metric,
using the PCA approach.
 The latter
 defines a metric that is intrinsic to the data, determined by
 their second order statistics. But even then, several
 possibilities exist, leading to non-equivalent results.

 Principal component decomposition can be applied both to the correlation
 matrix $C_{\alpha \beta}=<x_\alpha x_\beta>$
 and to the covariance matrix
 \be
  {\cal C}_{\alpha \beta}=<(x_\alpha -<x>_\alpha) (x_\beta -<x>_\beta)>
  =    C_{\alpha \beta} -  <x>_\alpha <x>_\beta\, .
 \ee
 In both cases averaging is performed over all data points, and the
  indices indicate spatial coordinates from 1 to $d$.
  The principal components are the eigenvectors of these matrices.
 Thus we have two natural bases in which to
  represent the data.
  Moreover one often renormalizes the eigenvector projections, dividing them
  by the square-roots of their eigenvalues. This procedure is known
  as ``whitening", leading to a renormalized correlation or covariance
  matrix of unity.
 This is a
 scale-free representation, that would naturally lead one to start
 with $\sigma=1$ in the search for (higher-order) structure of the data.

 The PCA approach that we have used on our examples was based on whitened
 correlation matrix projections.
  This turns out to produce good separation
 of crab-data in PC2-PC3 and of iris-data in PC1-PC2.
 Had we used the covariance matrix ${\bf \cal C}$ instead,
 we would get similar, but
 slightly worse, separation of crab-data, and a much worse representation
 of the iris-data.
 Our examples  are meant to convince the reader that
 once a good metric is found, QC conveys the correct
 information. Hence we allowed ourselves to search first for the best
 geometric representation, and then apply QC.

\section{ QC in Higher Dimensions}

In the iris problem we obtained excellent clustering results using the
first two principal components, whereas in the crabs problem, clustering
that depicts correctly the classification necessitates components 2 and 3.
However, once this is realized, it does not harm to add the 1st component.
This requires working in a 3-dimensional space, spanned by the three
leading PCs. Increasing dimensionality means higher
computational complexity, often limiting the applicability of a numerical
method. Nonetheless, here we can overcome this ``curse of dimensionality"
by limiting ourselves to evaluating $V$ at locations of data-points only. Since
we are interested in where the minima lie, and since invariably they lie
near data points, no harm is done by this limitation. The results are
depicted in Fig. 5. Shown here are $V/E$ values as function of the serial
number of the data, using the same symbols as in Fig. 2 to allow for
visual comparison. Using all data of $V<0.3E$ one obtains cluster cores
that are well separated in space, corresponding to the four classes that
exist in the data. Only 9 of the 129 points that obey $V<0.3E$ are
misclassified by this procedure.
Adding higher PCs, first component 4 and then component 5, leads to
deterioration in clustering quality. In particular, lower cutoffs in $V/E$,
including lower fractions of data, are required to define cluster cores
that are well separated in their relevant spaces.

One may locate the cluster centers, and deduce the clustering allocation
of the data, by following dynamics of gradient descent into the
potential minima. Defining $\y_i(0)=\x_i$ one follows steps of
$\y_i(t+\Delta t)=\y_i(t)-\eta(t) \nabla V(\y_i(t))$,
letting the points $\y_i$ reach an
asymptotic fixed value coinciding with a cluster center.
More sophisticated minimum search algorithms (see, e.g.
chapter 10 in \cite{press})
can be applied to reach the fixed points faster.
The results of a gradient-descent procedure,
applied to the 3D analysis of the crabs data
shown in Fig. 5, is displayed in Fig. 6. One clearly observes the four clusters,
and can compare clustering allocation with the original data classes.

\section{ Distance-based QC Formulation}

Gradient descent calls for the calculation
of $V$  both on the original data-points, as well as on the trajectories
they follow. An alternative approach can be to restrict oneself
to the original values of $V$, as in the example displayed in Figure 5,
and follow a hybrid algorithm to be described below. Before turning
to such an algorithm let us note that, in this case,
we evaluate $V$ on a discrete set of points
$V(\x_i)=V_i$. We can then express $V$ in terms of the distance matrix
$ D_{ij}=|\x_i - \x_j| $
as
\be
V_i=E-{d \over 2}+{1\over 2\sigma^2}{\sum_j D_{ij}^2
e^{- {D_{ij}^2 \over  2\sigma^2}} \over \sum_j e^{-{{D_{ij}^2}\over 2\sigma^2}}}
\ee
with $E$ chosen appropriately so that min$V_i$=0.
This kind of formulation is of particular importance if the original information
is given in terms of distances between data points rather than their locations
in space. In this case we have to proceed with distance information only.

Applying QC we can reach results of the type of Fig. 5 without invoking
any explicit spatial distribution of the points in question. One may then
analyze the results by choosing
 a cutoff, e.g. $V <0.2 E$, such that a fraction
(e.g. $1/3$) of the data will be included.
On this subset select groups of
points whose distances from one another are
smaller than, e.g., 2$\sigma$, thus defining cores of clusters.
Then continue with higher values of $V$, e.g. $0.2 E< V <0.4 E$, allocating
points to previous clusters or forming new cores. Since the choice of
distance cutoff in cluster allocation is quite arbitrary, this method cannot
be guaranteed to work as well as the gradient-descent approach.

\section{ Generalization}

Our method can be easily generalized to allow for different weighting of
different points, as in
\be
\psi(\x) =  \sum_i c_i e^{-{({\bf x -x}_i)^2 \over 2\sigma^2}}
\ee
with $c_i \geq 0$.
This is important if we have some prior information or some other means
for emphasizing or deemphasizing the influence of data points.
An example of the latter is using QC in conjunction with SVC \cite{nips00}.
SVC has
the possibility of labelling points as outliers. This is done
by applying quadratic maximization to the Lagrangian
\be \label{wolfe}
W=1 - \sum_{i,j} \beta_i \beta_j e^{-{(\x_i-\x_j)^2 \over 2\sigma^2}}
\ee
over the space of all $0 \leq \beta_i \leq {1 \over pN}$
subject to the constraint $\sum_i \beta_i =1$. The points for which the
upper bound of
$\beta_i$ is reached are labelled as outliers. Their number is
regulated by $p$, being limited by $pN$.
Using for the QC analysis a choice of $c_i={1 \over pN}-\beta_i$
 will eliminate the outliers of SVC and emphasize
the role of the points expected to lie within the clusters.

\section{ Discussion}

QC constructs a potential function $V(\x)$
on the basis of data-points, using one parameter, $\sigma$,
that controls the width of the structures that we search for.
The advantage of the potential $V$ over the scale-space probability
distribution is that the minima of the former are better defined
(deep and robust) than the maxima of the latter. However, both of
these methods put the emphasis on cluster centers, rather than, e.g.,
cluster boundaries.  Since the equipotentials of $V$ may take
arbitrary shapes, the clusters are not spherical, as in the k-means
approach. Nonethelss, spherical clusters appear more naturally
than, e.g., ring-shaped or toroidal clusters, even if the data would
accomodate them.
If some global symmetry is to be expected, e.g. global spherical symmetry,
it can be incorporated in the original Schr\"odinger
equation defining the potential function.

QC can be applied in high dimensions by limiting
the evaluation of the potential, given as an explicit
analytic expression of Gaussian terms, to locations
of data points only. Thus the complexity of evaluating $V_i$ is of order
$N^2$ independently of dimensionality.
Our algorithm has one free parameter, the scale $\sigma$. In all examples
we confined ourselves to scales that are of order 1, because we have
worked within whitened PCA spaces. If our method is applied
to a different data-space,
the range of scales to be searched for could be
determined by some other prior information.

Since the strength of our algorithm lies in the easy selection of cluster
cores, it can be used as a first stage of a  hybrid approach employing other
techniques after the identification of cluster centers.
The fact that we do not have to take care of feeble minima, but consider only
robust deep minima, turns the identification of a core into an easy problem.
Thus, an approach that drives its rationale from physical intuition in
quantum mechanics, can lead to interesting results in the field of
pattern classification.


We thank B. Reznik for a helpful discussion.


\newpage

\begin{figure}[]
 \centerline{\psfig{figure=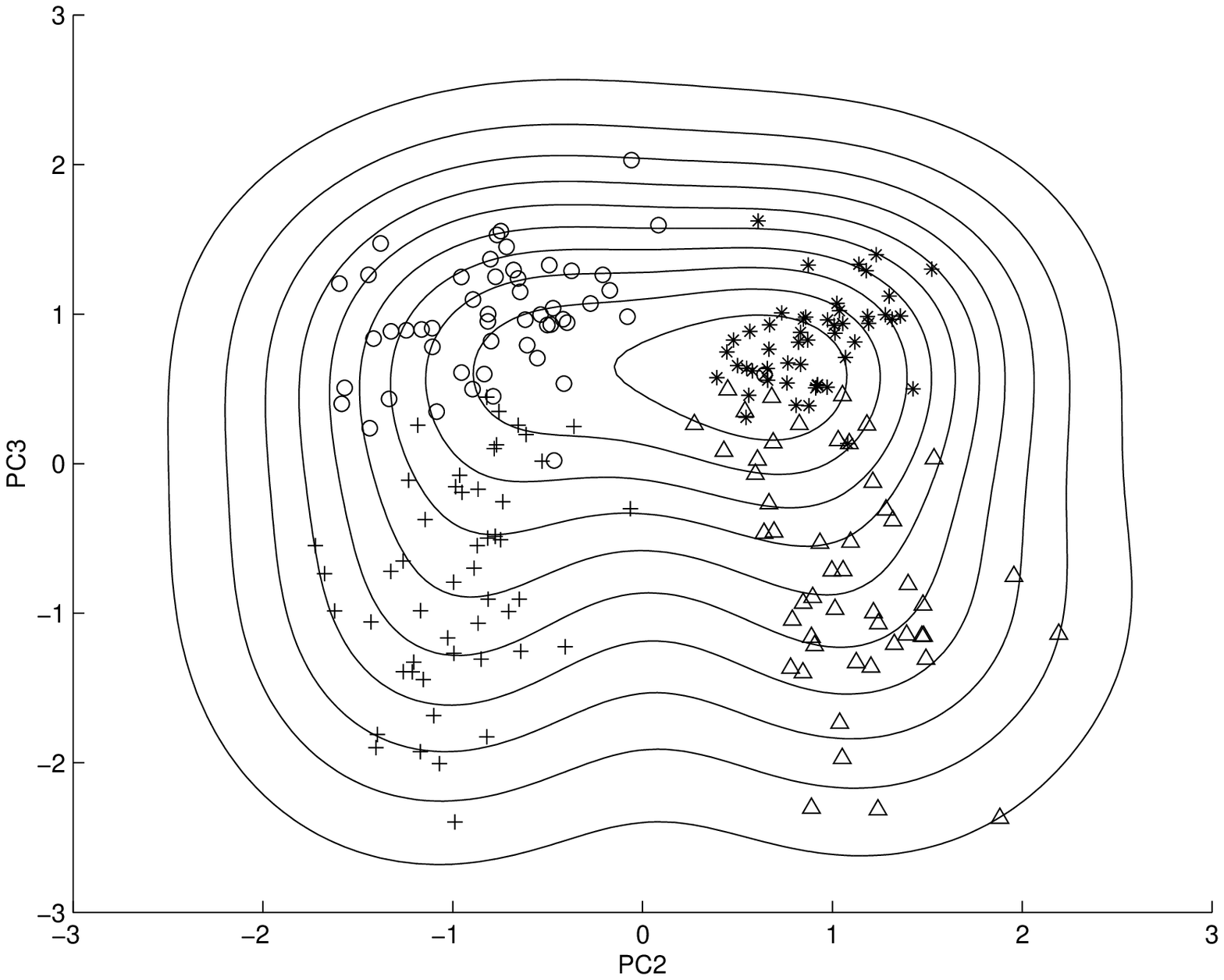,width=85mm}}
\vspace{2mm}
\caption{Ripley's crab data \cite{ripley} displayed on a plot of their
2nd and 3rd principal components with a superimposed topographic
map of the Roberts' probability distribution for $\sigma=1/\sqrt 2$.}
\label{Fig1}
\end{figure}

\begin{figure}[]
\centerline{\psfig{figure=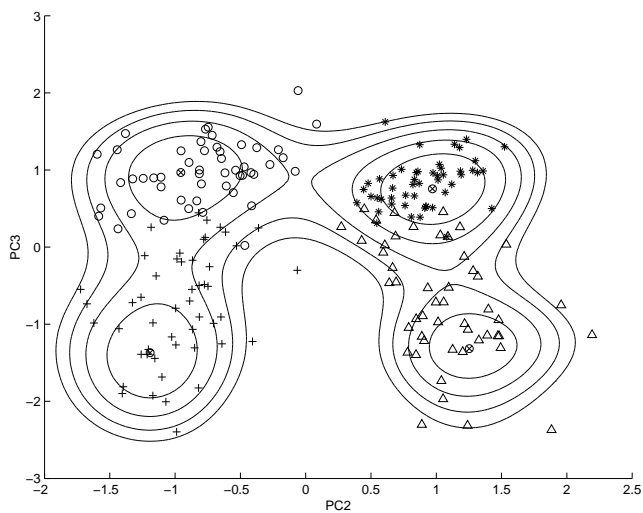,width=8.5cm}}
\caption{
A topographic map of the potential for the crab data with
$\sigma=1/\sqrt 2$,
displaying four minima (denoted by crossed circles)
that are interpreted as cluster centers.
The contours of the topographic map are set at values of
$V(\x)/E=.2,.4,.6,.8,1. $}
\end{figure}

\begin{figure}[]
\centerline{
\psfig{figure=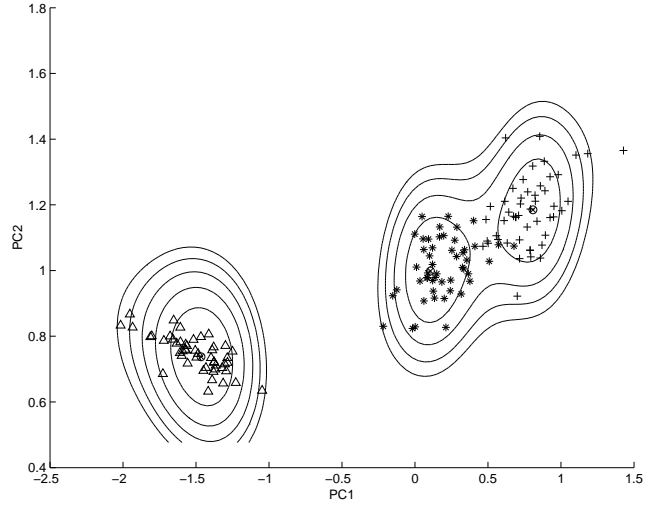,width=85mm}}
\caption{Quantum clustering of the iris data for $\sigma=1/4$ in a space
spanned by the first
two principal components. Different symbols represent the three
classes. Equipotential lines
are drawn at  $V(\x)/E=.2,.4,.6,.8,1.$  }
\end{figure}

\begin{figure}[]
\centerline{
\psfig{figure=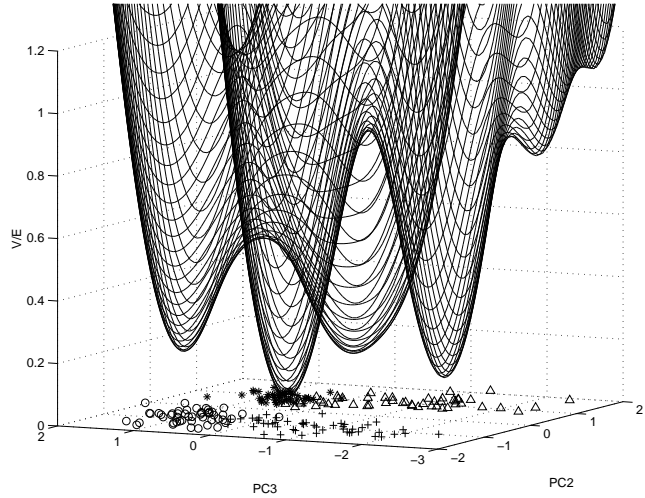,width=85mm}}
\caption{The potential for the crab data with $\sigma = 1/2$
 is displayed on a three-dimensional plot of $V(\x)$ in units of $E$,
the ground-state energy. Two additional, but insignificant, minima
appear.  The four deep minima are
roughly at the same locations as in Fig. 2.  }
\end{figure}

 \begin{figure}[]
\centerline{
\psfig{figure=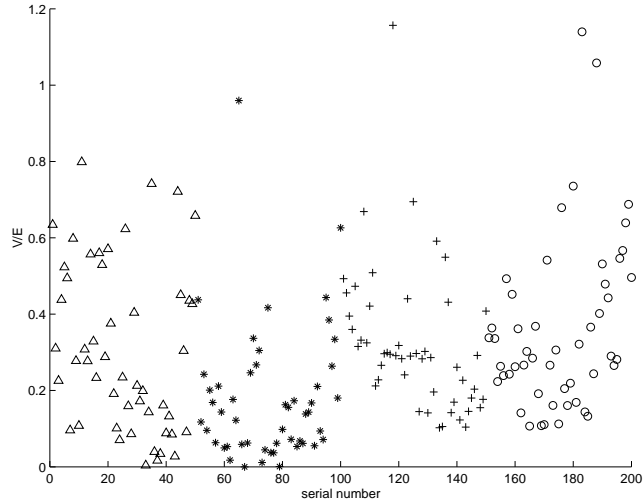,width=85mm}}
\caption{
 Values of $V(\x)/E$ are depicted in the crabs problem with
three leading PCs for $\sigma = 1/2$. They are presented as function
of the serial number of data, using the same symbols of data employed
before. One observes low lying data of all four classes.
}
\end{figure}

\begin{figure}[]
\centerline{
\psfig{figure=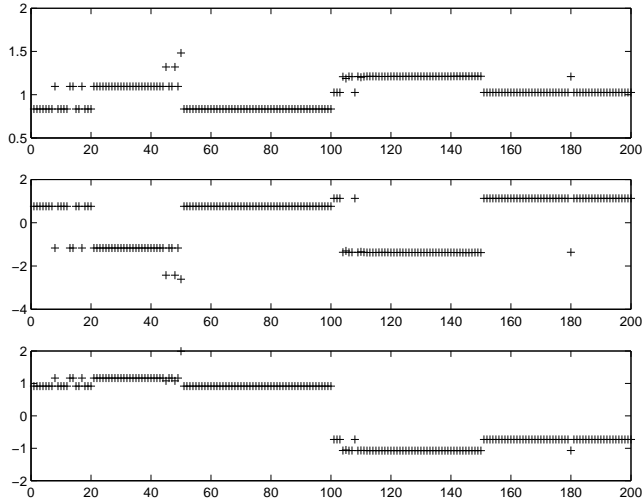,width=85mm}}
\caption{
 The results of gradient-descent dynamics applied to the 3D
 problem presented in Fig. 5. Shown here are values of the three
 coordinates (the three PCs) in three windows, as function
 of the serial number of the data. One clearly observes four clusters.
 But for the first class of data (the first 50 points), all clusters agree
 quite well with the classes. The first class, although belonging mostly
 to a separate cluster, has a large overlap with the cluster of the second class.
 Three points (numbers 45, 48 and 50) do not belong to any of
  the four clusters.
}
\end{figure}
\end{document}